\documentclass[11pt]{article}
\usepackage{amsmath,amssymb,color,graphics,epsfig}

\usepackage{amsfonts}

\newcommand{\be}{\begin{equation}}
\newcommand{\ee}{\end{equation}}
\newcommand{\bea}{\setlength\arraycolsep{2pt} \begin{eqnarray}}
\newcommand{\eea}{\end{eqnarray}}
\newcommand{\nn}{\nonumber}

\newcommand{\mm}{\mathrm}

\def\ft#1#2{{\textstyle{\frac{\scriptstyle #1}{\scriptstyle #2} } }}
\def\fft#1#2{{\frac{#1}{#2}}}

\def\0{{\sst{(0)}}}
\def\1{{\sst{(1)}}}
\def\2{{\sst{(2)}}}
\def\3{{\sst{(3)}}}
\def\4{{\sst{(4)}}}
\def\5{{\sst{(5)}}}
\def\6{{\sst{(6)}}}
\def\7{{\sst{(7)}}}
\def\8{{\sst{(8)}}}
\def\sst#1{{\scriptscriptstyle #1}}

\begin{document}

\begin{flushright}
\end{flushright}

\vspace{25pt}
\begin{center}
{\large {\bf Analytical approach to criticality of AdS black holes}}

\vspace{10pt}
 Hong-Ming Cui$^{1\dagger}$ and Zhong-Ying Fan$^{1\dagger}$

\vspace{10pt}
$^{1\dagger}${ Department of Astrophysics, School of Physics and Material Science, \\
 Guangzhou University, Guangzhou 510006, P.R. China }\\

\vspace{40pt}

\underline{ABSTRACT}
\end{center}
We establish a hidden symmetry between the specific volumes of the coexistent phases and hence develop an analytical approach to study criticality of AdS black holes. In particular, using the method, we solve the coexistence line exactly for a variety of black holes, including the charged AdS black hole in diverse dimensions, the rotating AdS black hole, the Gauss-Bonnet black hole and the quantum BTZ black hole as well as the Van der Waals fluid.

\vfill {\footnotesize  Email: fanzhy@gzhu.edu.cn\,.}

\thispagestyle{empty}

\pagebreak

\tableofcontents
\addtocontents{toc}{\protect\setcounter{tocdepth}{2}}




\section{Introduction}
In the past two decades, extended thermodynamics of Anti-de-Sitter (AdS) black holes was widely studied in the literature since the pioneer work \cite{Kubiznak:2012wp}, in which the cosmological constant was identified to a thermodynamic pressure and its conjugate as a thermodynamic volume. An advantage of such extension is the small-large  black hole transition of the charged AdS black hole (in canonical ensemble) looks similar to that of a Van der Waals fluid. This is very interesting in holography and has facilitated numerous developments in black hole physics, such as the black hole chemistry \cite{Karch:2015rpa}, the microstructures of black holes \cite{Wei:2015iwa,Wei:2019uqg} and new bounds on the black hole entropy \cite{Cvetic:2010jb,Amo:2023bbo}. Recently, the phase diagram of AdS black holes was extended to the coexistence region \cite{Wei:2023mxw} and the supercritical region \cite{Wang:2025ctk}. Actually the black hole thermodynamics can be further extended by varying all coupling constants. For instance, the central charge criticality was studied in \cite{Cong:2021fnf,Cong:2021jgb,Cui:2024cnj,Kumar:2022afq} by varying the Newton's gravitational constant and in \cite{Cui:2024xyl}, global monopole charge criticality was examined in diverse dimensions. 

However, in all these cases, the phase structure of AdS black holes was in general studied only numerically since the coexistence line cannot be solved exactly, except a few examples \cite{Cui:2024xyl,Spallucci:2013osa,Cui:2025qdy}. This is somehow disappointing since unlike the fluid (gas) in experiments, the equation of states (eos) for AdS black holes generally have analytical expressions. In this work, we will show that once the eos is given, an analytical approach to criticality of AdS black holes could be developed. This is possible because of a simple fact: the functional relation between the coexistent small ($r_s$)-large ($r_l$) black hole sizes is {\it self-reciprocal}. That is if $r_s=\varphi(r_l)$, then $r_l=\varphi(r_s)$ or equivalently $\varphi=\varphi^{-1}$ (we refer these functions to as self-reciprocal). As a  consequence, one can write $\varphi=\varphi(r_*)$, where $r_*$ collectively denotes the coexistent black hole sizes. Using this fact, we develop a simple approach, in which the small-large black hole sizes can be determined by a single algebraic equation for $\varphi(r_*)$ so that the coexistence line can be generally studied half-analytically at least. In particular, we will show that the coexistence line can be solved exactly for a variety of black holes in this approach, for example the charged AdS black hole in diverse dimensions, the rotating AdS black hole, the Gauss-Bonnet black hole, the quantum BTZ black hole  as well as the Van der Waals fluid.

The remainder of this paper is organized as follows. In section 2, we clarify the functional relation between the coexistent small-large black hole sizes and develop a general approach to study the coexistence line analytically. As an example, we adopt the method to a Van der Waals fluid. In section 3, we study the charged AdS black hole in diverse dimensions. The coexistence line is obtained exactly in general dimensions. In section 4, we study the leading corrections of angular momenta to a rotating AdS black hole. In this limit, we derive the coexistence line exactly.  In section 5, we study the P-V criticality of the Gauss-Bonnet black hole for a fixing higher order coupling constant $\alpha$. The coexistence line is solved exactly. In section 6, we study the $U-\nu$ criticality of the quantum BTZ black hole and reproduce the coexistence line reported in the literature.

\section{Self-reciprocal function and the coexistence line}

 Consider criticality of certain AdS black holes, for example a first order transition occurs between a pair of black holes below a critical temperature. The coexistence line is determined by
\be T(r_s)=T(r_l)\,,\qquad G(r_s)=G(r_l) \,,\label{mastercoe}\ee
where $T(r_h)$ stands for the temperature, $G(r_h)$ the Gibbs free energy and $r_s/r_l$ denotes the small/large black hole (horizon) size, respectively.
Generally the above equations can only be solved numerically, except a few examples \cite{Cui:2024xyl,Spallucci:2013osa,Cui:2025qdy}. However, we are aware of that the solution in fact can be found analytically according to a symmetry between the coexistent black hole sizes. To see this, we write 
\be r_l=\varphi(r_s)  \qquad \Rightarrow\qquad r_s=\varphi^{-1}(r_l) \,.\ee 
Substituting the relation into (\ref{mastercoe}) yields
\be T\big(r_* \big)=T\big( \varphi(r_*)\big)\,,\qquad  G\big(r_* \big)=G\big( \varphi(r_*)\big)\,,\label{coe1}\ee
and
\be T\big(r_* \big)=T\big( \varphi^{-1}(r_*)\big)\,,\qquad G\big(r_* \big)=G\big( \varphi^{-1}(r_*)\big) \,, \label{coe2}\ee
where $r_*$ stands for either the small black hole size $r_s$ or the large black hole size $r_l$. However, since there exists a pair of solutions for (\ref{mastercoe}), the equations Eq. (\ref{coe1}) and Eq. (\ref{coe2}) must have the same solutions, namely $r_*=r_s\,,r_l$ (this is why we have not written the subscripts explicitly in above equations). This implies that generally the function $\varphi$ should be {\it self-reciprocal}:  
\be \varphi(r_*)=\varphi^{-1}(r_*) \,.\ee 
That is the function is the same as its inverse function and vice versa. This illustrates a hidden symmetry between the coexistent phases. Explicitly we can write
\be r_s=\varphi(r_l)\qquad\mm{and}\qquad r_l=\varphi(r_s) \,.\ee
It is easy to see that once the function $\varphi$ is known, the coexistence line will be read off immediately (for example let $r_*$ runs from the small black hole side and then one can read off the corresponding large black hole size according to $r_l=\varphi(r_s)$). Here it should be emphasized that the above discussions are valid to a general quantum fluid, with the horizon size $r_h$ replaced by the specific volume $v$ of the fluid molecules. 

In fact, beyond the algebraic structure of the coexistence conditions, emergence of the self-reciprocal property could be attributed to spontaneous symmetry breaking of the system at the critical point. Consider a Ising-like model at first. Above the critical temperature, the system is in the paramagnetic phase and enjoys the Ising symmetry. At the critical point, the symmetry is spontaneously broken and  the system will be in either the positive-ferromagnetic phase (magnetization $M>0$) or the negative-ferromagnetic phase ($M<0$) below the critical temperature. The both are thermodynamically preferred on an equal footing and hence can coexist ( but no first order transition truly occurs between the two phases in the thermodynamic limit). The magnetization $M$ for the two phases exhibits a $Z_2$ symmetry: $M\rightarrow -M$. These results are standard in textbooks about critical phenomenon. Here our new observation is the emergent $Z_2$ symmetry between the ordered phases is a typical example of the self-reciprocal property, corresponding to $\varphi(M)=-M$.  As a comparison, the liquid-gas transition is not associated to the change of structural order. Despite the difference, the transition is still of second order at the critical point and hence  a certain symmetry of the supercritical fluid is spontaneously broken. This inspires us to interpret the self-reciprocal property  between the liquid-gas phases as a remnant of the spontaneous symmetry breaking. The story is much like the Ising-like systems. If this is correct, the self-reciprocal function $\varphi(v)$ specifies a symmetry for the microscopic theory of a supercritical fluid.


To proceed, let us consider a little math about the self-reciprocal function at first. The simplest example is $y=x$. However, this generally corresponds to trivial solutions (namely $r_s=r_l$) except at the critical point. For our purpose, a simple but sufficiently non-trivial example is the reciprocal function $y=c/x$, where $c$ is an arbitrarily positive constant. Actually this indeed gives the function $\varphi$ for certain examples (such as the Gauss-Bonnet black hole and the quantum BTZ black hole, see section 5 and 6 for details). However, generally the function $\varphi$ could be complicated. Nonetheless, we will adopt the following ansatz
\be r_l=\phi(r_s)/r_s \qquad \mm{or}\qquad r_s=\phi(r_l)/r_l \,.\label{masteransatz}\ee
 Notice that since $\varphi=\varphi(r_*)$, where $r_*$ runs from the coexistent small black hole size to the large black hole size, one has 
\be \phi=\phi(r_*)\,.\ee
However, unlike $\varphi$, this function is generally not self-reciprocal. Nevertheless, the above ansatz (\ref{masteransatz}) is extremely useful. We will show that by substituting it into the original problem (\ref{mastercoe}), the small-large black hole sizes will be determined by a single algebraic equation about the function $\phi(r_*)$.

\subsection{Van der Waals fluid}
Before moving to certain black holes, let us show how the ansatz (\ref{masteransatz}) works for a general quantum fluid. Here we consider a typical example:  the Van der Waals (VdW) fluid. The equation of state reads \cite{Kubiznak:2012wp} (we have set the Boltzmann constant $k_B=1$)
\be T=\Big( P+\fft{a}{v^2}\Big)(v-b) \,,\ee 
where $T$ is the temperature, $P$ the pressure and $v$ the specific volume of the fluid molecule. The constant $b>0$ describes the nonzero size corrections from the molecules whereas the constant $a>0$ is a measure of the attraction between them. The critical point can be found  from  the inflection point condition, given by \cite{Kubiznak:2012wp}
\be T_c=\fft{8a}{27b}\,,\quad v_c=3b\,,\quad P_c=\fft{a}{27b^2} \,.\ee
The Gibbs free energy reads
\be G=-T\left(1+\mm{ln}\Big[\fft{(v-b)T^{3/2}}{\Phi}\Big]\right)-\fft{a}{v}+P v \,,\ee
where $\Phi$ is a dimensionful constant characterizing the gas. Since the absolute value of $\Phi$ is unimportant, without loss of generality we will set $\Phi=1$. For simplicity, here and below, we work in normalized quantities $t=T/T_c\,,g=G/G_c$ and $z=v/v_c\,,p=P/P_c$. The normalized temperature and the Gibbs free energy are given by
\bea
&& t(z)=\fft{(3z-1)(3+p z^2)}{8z^2}\,,\nn\\
&& g(z)=\fft{\#}{z^2}\left[2pz^2-36z+6-(3z-1)(3+pz^2)\mm{ln}\Big( \fft{a^3(3z-1)^5(3+pz^2)^3}{27^3b z^6} \Big)\right]\,,
\eea
where $\#$ stands for an unimportant numerical factor.
\begin{figure}
  \centering
  \includegraphics[width=270pt]{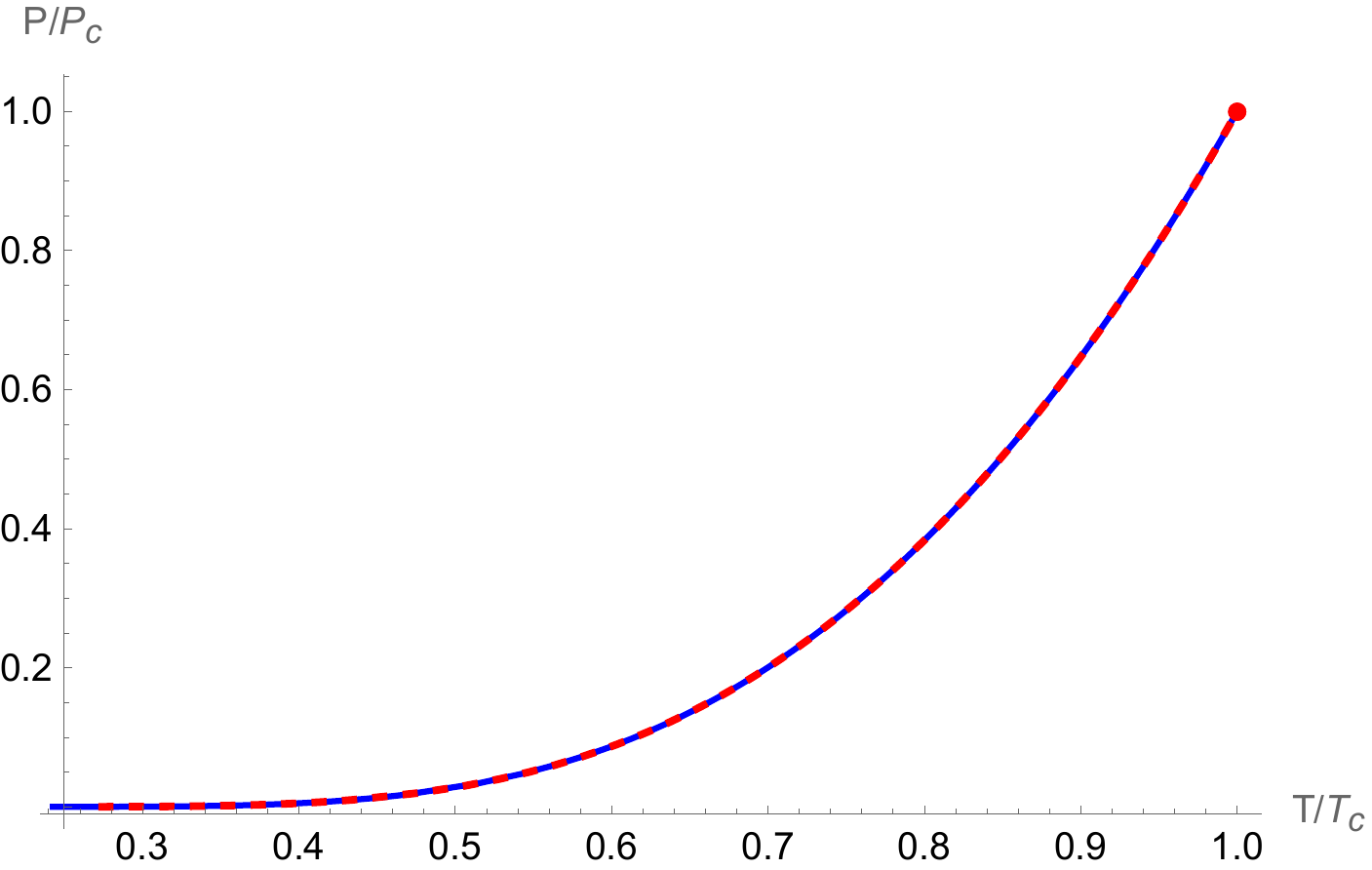}
  \caption{The coexistence curve on the $T-P$ plane for the VdW fluid. The solid line is the analytical solution whereas the dashed line stands for the numerical solution.}
\label{VdW}
\end{figure}
To solve the coexistence curve, we substitute the ansatz 
\be z_l=\phi(z_s)/z_s \qquad \mm{or}\qquad z_s=\phi(z_l)/z_l \,,\label{masteransatz1}\ee
into the equations 
\be t(z_s)=t(z_l)\,,\qquad g(z_s)=g(z_l) \,.\label{mastereq}\ee
From the first equation, we solve the pressure 
\be p_*=\fft{(3z_*-1)\phi-z_*^2}{\phi^2z_*} \,.\label{vdwp}\ee
Notice that since $\phi$ is positive definite, $p_*>0$ requires $z_*>1/3$, which is a lower bound for the specific volume.
By plugging (\ref{vdwp}) into the second equation, we deduce
\be 6(\phi-z_*^2)\Big( (6z_*-1)\phi-z_*^2 \Big)+2(3\phi-z_*)(3z_*-1)(\phi+z_*^2)\,\mm{ln}\left[ \fft{z_* (3z_*-1)}{(3\phi-z_*)} \right]=0 \,.\label{vdwmaster}\ee
Here $3\phi-z_*>0$ owing to $p_*>0$. This is a single algebraic equation determining the function $\phi(z_*)$. Once $\phi(z_*)$ is known, the coexistent temperature can be read off directly from
\be t_*=\fft{(3z_*-1)(3+p_* z_*^2)}{8z_*^2} \,. \label{vdwt}\ee
In fact, this has already given the coexistence curve on the $T-v$ plane since $z=v/v_c$. Moreover, combining the relations (\ref{vdwp}) and (\ref{vdwt}), one obtains the coexistence curve on the $T-P$ plane. This specifies a general procedure to solve the coexistence line half-analytically. 

While for the VdW fluid, the function $\phi(z_*)$ cannot be solved directly (owing to the logarithmic function), analytical solution to the coexistence line can be obtained by introducing a new parameter 
\be x\equiv \fft{z_*(3z_*-1)}{3\phi-z_*}   \,.\ee
Since $\phi=z_s z_l$, this parameter describes either the ratio $(3z_s-1)/(3z_l-1)$ (when $z_*=z_s$) or its inverse (when $z_*=z_l$). In other word, $x\leq 1$ ($x\geq 1$) describes the liquid (gas) phase, respectively (this can be checked from (\ref{zstarvdw})). It turns out that using $x$, $z_*$ can be solved explicitly from Eq. (\ref{vdwmaster})
\be z_*=\fft{(x-1)(x-\mm{ln}x-1)}{3(x+1)\mm{ln}x-6(x-1)} \,,\label{zstarvdw}\ee
which leads to
\bea
&&t_*=\fft{27\Big((x+1)\mm{ln}x-2(x-1)\Big)\big(x^2-2x\,\mm{ln}x-1\big)^2}{8(x-1)\big(x-\mm{ln}x-1\big)^2\big(x\,\mm{ln}x-x+1\big)^2}\,,\nn\\
&&p_*=\fft{27x\Big((x-1)^2-x\,\mm{ln}^2x\Big)\Big((x+1)\mm{ln}x-2(x-1)\Big)^2}{(x-1)^2\big(x-\mm{ln}x-1\big)^2\big(x\,\mm{ln}x-x+1\big)^2}\,.
\eea
Here the coexistence line can be read off analytically from either the liquid phase $x\leq 1$ or the gas phase $x\geq 1$. As depicted in Fig. \ref{VdW}, our analytical solution is perfectly matched with the ordinary numerical solution obtained by solving (\ref{mastereq}) directly. 

To end this section, we point out that our method is valid to a general quantum fluid exhibiting critical phenomenon as long as the equation of state and the Gibbs free energy are known analytically. This is of course the case for the holographic fluids dual to AdS black holes.

\section{Charged AdS black hole in diverse dimensions}
Consider the charged AdS black hole in general $D=d+3$ dimensional AdS spacetimes \cite{Cai:1998vy,Chamblin:1999tk}
\bea 
&&ds^2=-f(r)dt^2+\fft{dr^2}{f(r)}+r^2 d\Omega_{d+1}^2 \,,\quad A=\fft{16\pi }{d\omega_{d+1}}\left(\fft{Q}{r_h^d}-\fft{Q}{r^d}\right)dt \,,\nn\\
&&f(r)=\fft{r^2}{\ell^2}+1-\fft{16\pi M}{(d+1)\omega_{d+1}r^d}+\fft{128\pi^2 Q^2}{d(d+1)\omega^2_{d+1}r^{2d}}\,,
\eea
where $\omega_{d+1}$ stands for the volume of a unit $(d+1)$-dimensional sphere, $r_h$ the horizon radius, $\ell$ the AdS radius and $Q$ the electric charge carried by the black hole. The mass, the temperature, the entropy and the electrostatic potential are given by
\bea
&& M=\fft{\omega_{d+1}}{16\pi d}\left( d(d+1)r_h^d(1+r_h^2\ell^{-2})+\fft{128\pi^2Q^2}{\omega^2_{d+1}\,r_h^d} \right)\,,\nn\\
&& T=\fft{1}{4\pi r_h}\left( d+(d+2)r_h^2\ell^{-2}-\fft{128\pi^2Q^2}{(d+1)\omega^2_{d+1}r_h^{2d}} \right) \,,\nn\\
&& S=\fft{1}{4}\omega_{d+1} r_h^{d+1}\,,\qquad \Phi=\fft{16\pi Q}{d\omega_{d+1} r_h^{d}}\,,
\eea
Moreover, by identifying the cosmological constant to the thermodynamic pressure
\be  P=-\fft{\Lambda}{8\pi}\,,\qquad \Lambda=-\fft{(d+1)(d+2)}{2\ell^2} \,,\ee
the thermodynamic first law could be extended to \cite{Kubiznak:2012wp,Gunasekaran:2012dq}
\be dM=TdS+\Phi dQ+V dP \,, \ee
where the conjugate volume is evaluated to be 
\be V=\fft{\omega_{d+1}r_h^{d+2}}{d+2} \,.\ee
A remarkable feature of such extension is in canonical ensemble the small-large black hole transition is quite similar to that of a VdW fluid. The critical point occurs at \cite{Gunasekaran:2012dq}
\bea  
&& r_{c}^d=\fft{8\sqrt{2(2d+1)}\,\pi Q}{\sqrt{d}\,\omega_{d+1}}\,,\nn\\
&&P_c^d=\fft{d^{2d+1}\,\omega^2_{d+1}}{(2d+1)2^{7+4d}\,\pi^{2+d}Q^2}\,,\nn\\
&&T_c^d=\fft{d^{2d+\fft12}\,\omega_{d+1}}{8\sqrt{2}(2d+1)^{d+\fft12}\, \pi^{d+1}Q}\,,
\eea
where $r_c$ stands for the critical horizon radius. However, the coexistence line was only studied numerically in \cite{Kubiznak:2012wp,Gunasekaran:2012dq}. Later it was shown that in the four dimension, the coexistence line can be solved exactly by using the Maxwell's area law \cite{Spallucci:2013osa}. 

Here we will show that the coexistence line can be solved exactly in both the four and the five dimensions and can be analytically studied in general higher dimensions. Again we work with normalized quantities $z=r_h/r_c\,,t=T/T_c\,,p=P/P_c$. The normalized temperature and the Gibbs free energy $g=G/G_c$ are given by
\bea
&& t(z)=\fft{(d+1)(2d+1)z^{2d}-1+d(2d+1)p\, z^{2d+2} }{4d(d+1)z^{2d+1}}\,,\nn\\
&&g(z)=\fft{(d+1)\big( (d+1)z^{2d}+1 \big)-d^2p\,z^{2d+2}}{4(d+1)z^d}\,.
\eea
Then substituting the ansatz (\ref{masteransatz1}) into the equations (\ref{mastereq}), we solve the pressure at first 
\be p_*=\fft{(d+2)z_*^2 (\phi^d-z_*^{2d})\big((d+1)\phi^d-1 \big)}{d^2\phi^d\big( \phi^{d+2}-z_*^{2d+4} \big)} \,,\ee
and the original problem (\ref{mastereq}) reduces to 
\bea\label{RNeq} 
&& d\Big(  (d+1)(2d+1)z_*^{2d}-1 \Big)\phi^{3d+3}-2(d+1)^2(2d+1)z_*^{2d+2}\phi^{3d+2} \nn\\
&&+(d+1)(d+2)(2d+1)z_*^{2d+4}\phi^{3d+1}+(d+2)(2d+1)z_*^{2d+2} \Big( (d+1)z_*^{2d}+1 \Big)\phi^{2d+2}\nn\\
&&-2(d+1)^2z_*^{2d+4}\Big( (2d+1)z_*^{2d}+1 \Big)\phi^{2d+1} +d(d+1)(2d+1)z_*^{4d+6}\phi^{2d}\nn\\
&&-2(d+1)^2z_*^{4d+2}\phi^{d+2}+(d+2)(2d+1)z_*^{4d+4}\phi^{d+1}-d z_*^{6d+6}=0\,,      
\eea
which is a single algebraic equation about $\phi(z_*)$. Clearly once the function $\phi(z_*)$ is solved, the coexistent pressure and temperature will be extracted immediately since
\be  t_*=\fft{(d+1)(2d+1)z_*^{2d}-1+d(2d+1)p_*\, z_*^{2d+2} }{4d(d+1)z_*^{2d+1}}\,. \ee
This gives rise to an analytical result for the coexistence line in general dimensions. While the equation (\ref{RNeq}) looks quite complicated, it does lead to exact solutions in diverse dimensions.

\subsection{The four dimension}
In the four dimension $d=1$, the equation (\ref{RNeq}) simplifies to
\be (\phi-z_*^2)^4\Big( (6z_*^2-1)\phi^2-4z_*^2 \phi-z_*^4 \Big)=0  \,.\ee
The solution $\phi=z_*^2$ is trivial because of $z_s=z_l$. The nontrivial solution is determined by the second bracket, given by
\be \phi(z_*)=\fft{z_*^2\big(2+\sqrt{6z_*^2+3} \,\big)}{6z_*^2-1}   \,.\ee
As an example, it is straightforward to verify that if $z_l=\phi(z_s)/z_s$, then $z_s=\phi(z_l)/z_l$ (but the function $\phi$ itself is not self-reciprocal). That is the functional relation between the small-large black hole sizes is self-reciprocal, as expected.

It follows that the coexistent pressure and the temperature read
\bea
&& p_*=\fft{\big( \sqrt{6z_*^2+3}-2\big)^2}{z_*^4}\,,\nn\\
&& t_*=\fft{6z_*^2+5-3\sqrt{6z_*^2+3}}{2z_*^3} \,.
\eea
Notice that the latter has already given a close formula for the coexistence curve on the $T-V $ plane because of $z=(V/V_c)^{1/3}$. Moreover, by converting the black hole sizes in terms of the pressure according to
\bea
&& z_s^2\,z_l^2=\fft{1}{p_*} \,,\nn\\
&& z_s^2+z_l^2=\fft{2\big( 3-2\sqrt{p_*} \big)}{p_*} \,,
\eea
and using the fact $t_*=\big( t(z_s)+t(z_l) \big)/2$, we arrive at a compact result on the $T-P$ plane
\be t_*=\sqrt{ \fft{p_*( 3-\sqrt{p_*} )}{2} }  \,.\label{d4tp}\ee 
This reproduces the result firstly obtained in \cite{Spallucci:2013osa}. However, the self-reciprocal property between the small-large black holes was unaware in that paper and the method there is limited to the four dimension. Below we will show that self-reciprocal property enables us to analytically solve the coexistence line for the charged black hole in diverse dimensions.

\subsection{The five dimension}
In the five dimension, the equation (\ref{RNeq}) reduces to 
\be 2(\phi+z_*^2)(\phi-z_*^2)^4\Big( (15z_*^4-1)\phi^4 -3z_*^2 \phi^3-7z_*^4 \phi^2-3z_*^6 \phi-z_*^8 \Big)=0  \,.\ee
Clearly the nontrivial solution is determined by the large bracket. One finds
\be \phi(z_*)=\fft{1}{12}\left( \fft{9z_*^2}{15z_*^4-1}+\sqrt{\Delta_1 }+\sqrt{\Delta_2} \right) \,, \ee
where
\bea
&& \Delta_1=\fft{3z_*^4}{15z_*^4-1}\left[ \fft{840z_*^4-29}{15z_*^4-1}-16(90z_*^4-17)\Sigma^{-1/3}+2\Sigma^{1/3} \right]\,, \\
&&\Delta_2=\fft{6z_*^4}{15z_*^4-1}\left[ \fft{840z_*^4-29}{15z_*^4-1}+8(90z_*^4-17)\Sigma^{-1/3}-\Sigma^{1/3}+\fft{27z_*^2\big( 1800z_*^8+180z_*^4-11\big)}{(15z_*^4-1)^2\sqrt{\Delta_1}} \right] \,,  \nn\eea
where
\be \Sigma=4\left[-3915z_*^4-101+9\sqrt{15\big(19200z_*^{12}+1735z_*^8+2706z_*^4-121 \big)} \right]\,.
\ee
The coexistent pressure and the temperature are given by
\bea\label{d5tp}
&& p_*=\fft{(3\phi^2-1)z_*^2}{\phi^2(\phi^2+z_*^4)} \,,\nn\\
&& t_*=\fft{15z_*^4-1+10p_* z_*^6}{24z_*^5} \,.
\eea
\begin{figure}
  \centering
  \includegraphics[width=210pt]{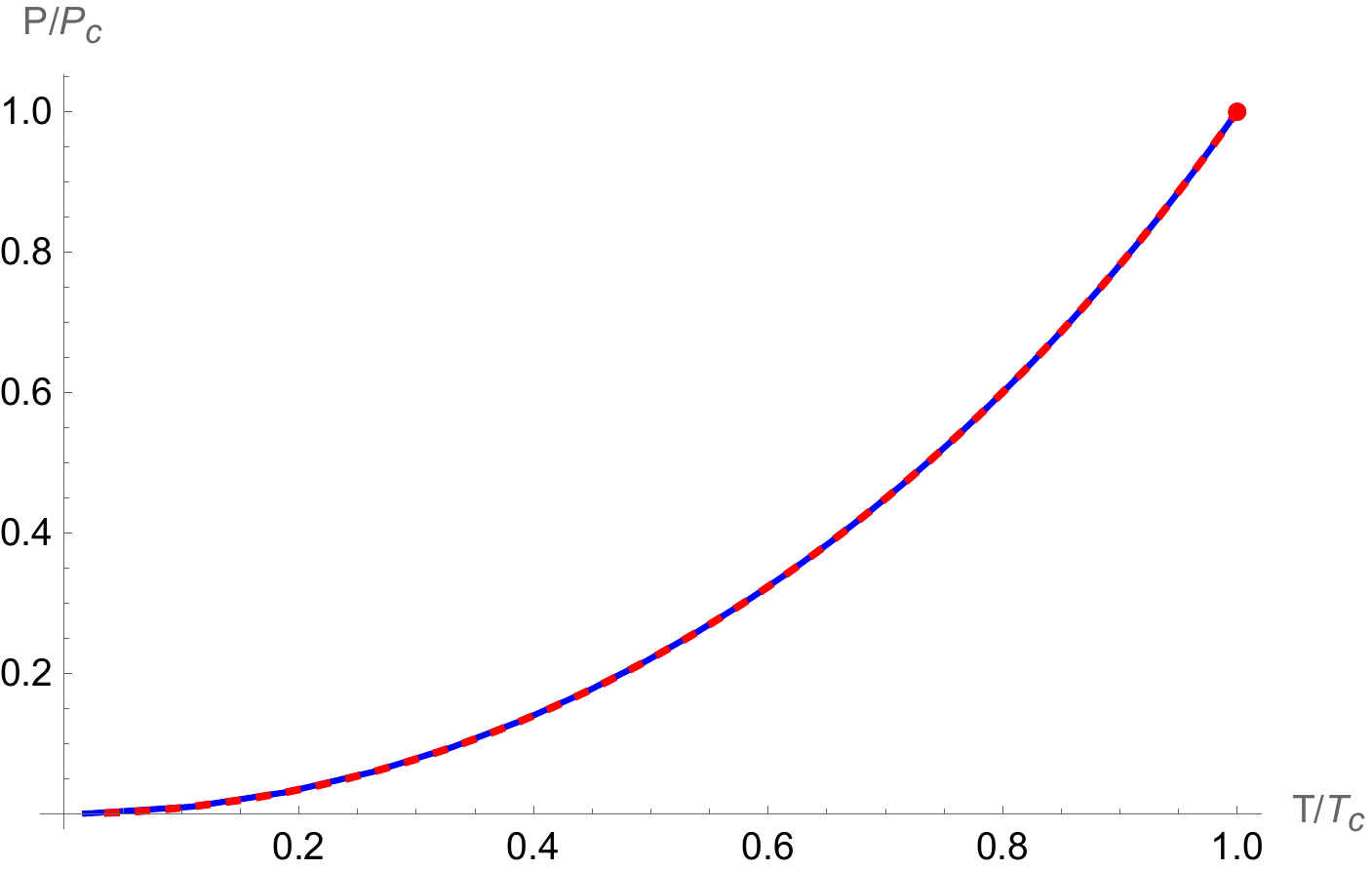}
  \includegraphics[width=210pt]{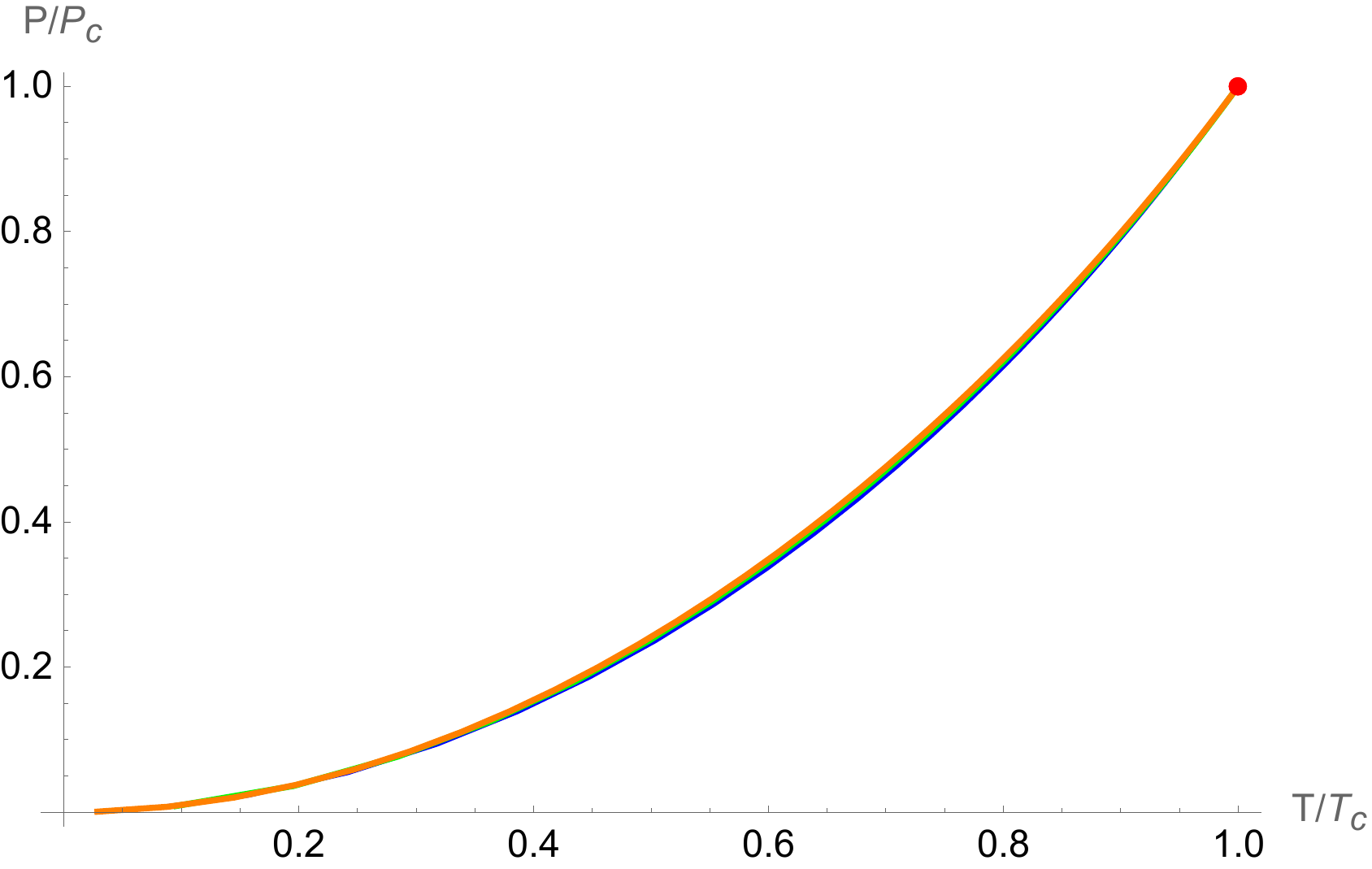}
  \caption{Left: the coexistence curve for the five dimensional charged black hole. The solid line is the analytical solution whereas the dashed line stands for the numerical solution. Right: the analytical coexistence curve for the $D=6$ (blue), $D=7$ (green) and $D=8$ (orange) dimensional charged black hole, respectively. }
\label{RNPV5}
\end{figure}
This gives the coexistence curve on the $T-P$ plane analytically, which is perfectly matched with the numerical result, as shown in the left panel of Fig. \ref{RNPV5}.

\subsection{Diverse dimensions}
However, in $D\geq 6$ higher dimensions, the function $\phi(z_*)$ cannot be solved directly  any longer. Nevertheless, analytical solution to the coexistence line can be obtained by introducing a new parameter 
\be x\equiv z_*^2/\phi \,.\ee
Notice that $x=z_s/z_l$ when $z_*=z_s$ or $x=z_l/z_s$ when $z_*=z_l$. This implies that $x\leq 1$ ($x\geq 1$) describes the small (large) black hole phase, respectively. Remarkably, using this parameter, Eq. (\ref{RNeq}) reduces to a linear equation for $z_*^{2d}$, which can be solved immediately as
\be z_*^{2d}=\ft{d\big( x^{3d+3}-5x^{2d+2}+4x^{2d+1}+4x^{d+2}-5x^{d+1}+1 \big)-2(d^2+1)(x-1)(x^d-1)x^{d+1}}{(d+1)(2d+1)x^{2d}(x-1)\Big( d x^{d+2}-(d+2)x^{d+1}+(d+2)x-d \Big)}  \,.\ee
We deduce
\bea
&& t_*(x)=\fft{(d+1)(2d+1)z_*^{2d}(x)-1+d(2d+1)p_*(x)\, z_*^{2d+2}(x) }{4d(d+1)z_*^{2d+1}(x)}\,,\nn\\
&& p_*(x)=\fft{ (d+2)x^2(x^d-1)\Big( (d+1)x^d z_*^{2d}(x)-x^d \Big) }{d^2 (x^{d+2}-1)z_*^{2d+2}(x)} \,.
\eea
The coexistence line on the $T-P$ plane can be read off from either the small black hole phase ($x\leq 1$) or the large black hole phase ($x\geq 1$). The solution is valid to diverse dimensions. In lower dimensions, one has for $D=4$ dimension 
\bea 
&& t_*=\fft{3\sqrt{6} x(x+1) }{(x^2+4x+1)^{3/2}}\,,\nn\\
&& p_*=\fft{36x^2}{(x^2+4x+1)^2} \,,
\eea
and for $D=5$ dimension
\bea
&& t_*=\fft{15^{5/4} x(x+1)^3 }{8(x^4+3x^3+7x^2+3x+1)^{5/4}} \,,\nn\\
&& p_*=\fft{3\sqrt{15} x^2(x^2+3x+1) }{ (x^4+3x^3+7x^2+3x+1)^{3/2}} \,.
\eea
It is straightforward to verify that these solutions are matched well with previous ones, given in (\ref{d4tp}) and (\ref{d5tp}). In $D\geq 6$ higher dimensions, the results (for several examples) are depicted in the right panel of Fig. \ref{RNPV5}. As far as we can check, our analytical solutions are matched with the numerical results in higher dimensions perfectly.

\section{Rotating AdS black hole}

Consider the Kerr-AdS black hole in the $D=4$ dimensions (under the Boyer-Lindquist coordinates)
\be ds^2=\rho^2\Big(\fft{dr^2}{\Delta_r}+\fft{d\theta^2}{\Delta_\theta} \Big)-\fft{\Delta_r}{\rho^2}\Big(dt-a\sin^2\theta \fft{d\phi}{\Xi} \Big)^2+\fft{\Delta_\theta\sin^2\theta}{\rho^2}\Big( adt-(r^2+a^2)\fft{d\phi}{\Xi} \Big)^2 \,,\label{kerr}\ee
where
\bea
&&\rho^2=r^2+a^2\cos^2\theta\,,\nn\\
&&\Delta_r=(r^2+a^2)(1+g^2r^2)-2M\Xi^2r\,,\nn\\
&&\Delta_\theta=1-g^2a^2\cos^2\theta\,,\quad \Xi=1-g^2a^2\,,
\eea
where $g=1/\ell$. The various thermodynamic quantities are given by
\bea
&& M=\fft{(r_h^2+a^2)(1+g^2a^2)}{2\Xi^2 r_h}\,,\nn\\
&& T=\fft{(3g^2r_h^2+1)r_h^2+(g^2r_h^2-1)a^2}{4\pi r_h(r_h^2+a^2)}\,,\quad S= \fft{\pi(r_h^2+a^2)}{\Xi}\,,\nn\\ 
&& J=Ma\,,\qquad \Omega=\fft{(1+g^2r_h^2)a}{r_h^2+a^2}\,,\nn\\
&& V=\fft{2\pi(r_h^2+a^2)\left(2r_h^2-(g^2r_h^2-1)a^2\right)}{3\Xi^2 r_h} \,.
\eea
where $J$ is the angular momenta, $\Omega$ the angular velocity and $V$ the thermodynamic volume. The first law is extended to $dM=TdS+\Omega dJ+VdP$. 

To study critical phenomenon of the solution, we shall consider the leading corrections from the angular momenta. Define the specific volume of the black hole \cite{Gunasekaran:2012dq}
\be v\equiv 2\left( \fft{3V}{4\pi} \right)^{1/3} \,.\ee 
The equation of state simplifies to
\be T=Pv+\fft{1}{2\pi v}-\fft{48J^2}{\pi v^5}+O(J^4) \,.\ee
In this approximation, the critical point occurs at
\be v_c=2\times 90^{1/4}\sqrt{J}\,,\quad T_c=\fft{90^{3/4}}{225\pi\sqrt{J}}\,,\quad P_c=\fft{1}{12\sqrt{90}\,\pi J} \,.\ee
\begin{figure}
  \centering
  \includegraphics[width=270pt]{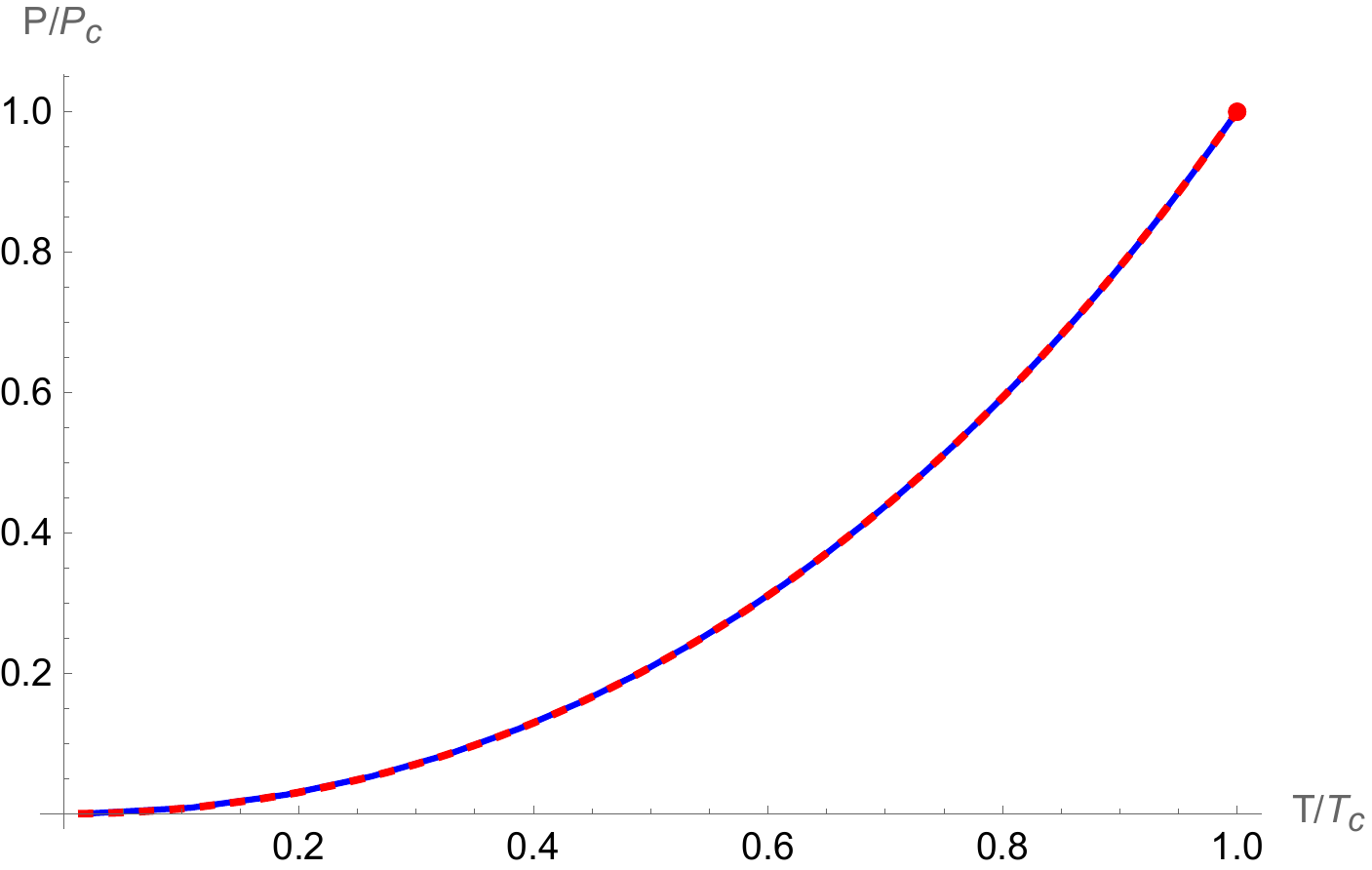}
  \caption{The coexistence curve for the rotating AdS black hole. The solid line is the analytical solution whereas the dashed line stands for the numerical solution.  }
\label{rotatingbh}
\end{figure}
The Gibbs free energy in the same approximation reads
\be G=\fft{v}{8}-\fft{\pi P v^3}{12}+\fft{20J^2}{v^3}  \,.\ee
The normalized temperature and the Gibbs free energy are given by
\bea
&& t=\fft{15z^4-1+10pz^6}{24z^5} \,,\nn\\
&& g=\fft{9z^4+1-2p z^6}{8z^3} \,,
\eea
where $z=v/v_c$. Again by plugging the ansatz (\ref{masteransatz1}) into the equations (\ref{mastereq}), we find
\be p_*=\fft{ (15z_*^4-1)\phi^4-z_*^2\phi^3-z_*^4 \phi^2-z_*^6\phi-z_*^8 }{10\phi^5z_*^4} \,,\ee
where the function $\phi$ is determined by a quartic algebraic equation
\be\label{masterkerr} (15z_*^4-1)\phi^4-4z_*^2\phi^3-5z_*^4 \phi^2-4z_*^6\phi-z_*^8 =0 \,.\ee 
The physical solution can be solved exactly as 
\be \phi(z_*)=\fft{1}{6}\left( \fft{6z_*^2}{15z_*^4-1}+\sqrt{\Delta_1 }+\sqrt{\Delta_2} \right) \,, \ee
where
\bea
&& \Delta_1=\fft{3z_*^4}{15z_*^4-1}\left[ \fft{2(75z_*^4+1)}{15z_*^4-1}-(180z_*^4+11)\Sigma^{-1/3}+\Sigma^{1/3} \right]\,, \\
&&\Delta_2=\fft{3z_*^4}{15z_*^4-1}\left[ \fft{4(75z_*^4+1)}{15z_*^4-1}+(180z_*^4+11)\Sigma^{-1/3}-\Sigma^{1/3}+\fft{36z_*^2\big( 450z_*^8+15z_*^4+1\big)}{(15z_*^4-1)^2\sqrt{\Delta_1}} \right] \,,  \nn\eea
where
\be 
\Sigma=540z_*^4-17+18\sqrt{5\big(3600z_*^{12}+840z_*^8+29z_*^4+1 \big)} \,.
\ee
The coexistent temperature can be read off as
\be t_*=\fft{15z_*^4-1+10p_*z_*^6}{24z_*^5} \,.\ee
As depicted in Fig. \ref{rotatingbh}, our analytical solution is perfectly matched with the numerical result reported in \cite{Gunasekaran:2012dq}.

In fact, using a different parametrization, the solution to the coexistence line can be written even more compactly. Define $x=z_*^2/\phi$. Eq. (\ref{masterkerr})  reduces to a linear equation for $z_*^4$, which can be solved as
\be z_*=\fft{(x^4+4x^3+5x^2+4x+1)^{1/4}}{15^{1/4}} \,.\ee
This gives rise to
\bea
&& t_*=\fft{5x(x+1)}{2(x^2+3x+1) z_*(x)}\,, \nn\\
&& p_*=\fft{3x^2(3x^2+4x+3)}{2(x^4+4x^3+5x^2+4x+1)z_*^2(x)}\,.
\eea
The coexistence line can be read off from either the small black hole phase ($x\leq 1$) or the large black hole phase ($x\geq 1$).

\section{Gauss-Bonnet black hole}
Consider the Gauss-Bonnet black hole in the five dimension  
\bea
&&ds^2=-f(r)dt^2+dr^2/f(r)+r^2 d\Omega^2_3\,, \nn\\
&&f(r)=1+\fft{r^2}{4\alpha}\Big[1-\sqrt{1-8\alpha\ell^{-2}+\ft{64\alpha G M}{3\pi r^4}}\, \Big] \,,
\eea
where $\ell$ stands for the bare AdS radius and $\alpha$ is the Gauss-Bonnet coupling, having dimension of length square. In holography, causality of the boundary theory constrains $0<\alpha<9\ell^2/200$ \cite{Brigante:2007nu,Camanho:2009vw}. The event horizon is defined by the largest real root of the equation $f(r_h)=0$. Using standard method, the mass, the entropy and the temperature can be evaluated as
\bea
&&M=\fft{3\pi( r_h^4+r_h^2\ell^2+2\alpha\ell^2 )}{8\ell^2}\,,\\
&&S=\fft{\pi^2 r_h^3}{2}\Big(1+\fft{12\alpha }{r_h^2} \Big) \,,\\
&&T=\fft{r_h(2r_h^2+\ell^2) }{2\pi \ell^2 (r_h^2+4\alpha )}\,.\label{tem}
\eea
Moreover, by identifying the bare cosmological constant to the thermodynamic pressure $P=6\ell^{-2}/8\pi$ and varying the Gauss-Bonnet coupling, the first law of thermodynamics can be extended to \cite{Cai:2013qga}
\be dM=TdS+VdP+Ud\alpha \,,\ee
where the thermodynamic volume and the chemical potential conjugate to $\alpha$ are given by
\be V=\fft{\pi^2r_h^4}{2}\,,\qquad U=-\fft{3\pi \big( 8r_h^4+3r_h^2\ell^2-4\alpha\ell^2 \big)}{4\ell^2(r_h^2+4\alpha)} \,.\ee 
The P-V criticality of the solution for a fixing $\alpha$ was previously numerically studied in \cite{Cai:2013qga}. Here we shall derive the coexistence line analytically. The critical point occurs at
\be r_c=2\sqrt{3\alpha}\,,\quad T_c=\fft{1}{4\pi\sqrt{3\alpha}}\,,\quad P_c=\fft{1}{96\pi\alpha} \,.\ee
The normalized temperature and the Gibbs free energy are given by
\bea
&& t=\fft{z(3+pz^2)}{3z^2+1}   \,,\nn\\
&& g=\fft{-6z^4+3z^2-1+pz^4(z^2+3)}{3z^2+1}    \,.
\eea
\begin{figure}
  \centering
  \includegraphics[width=270pt]{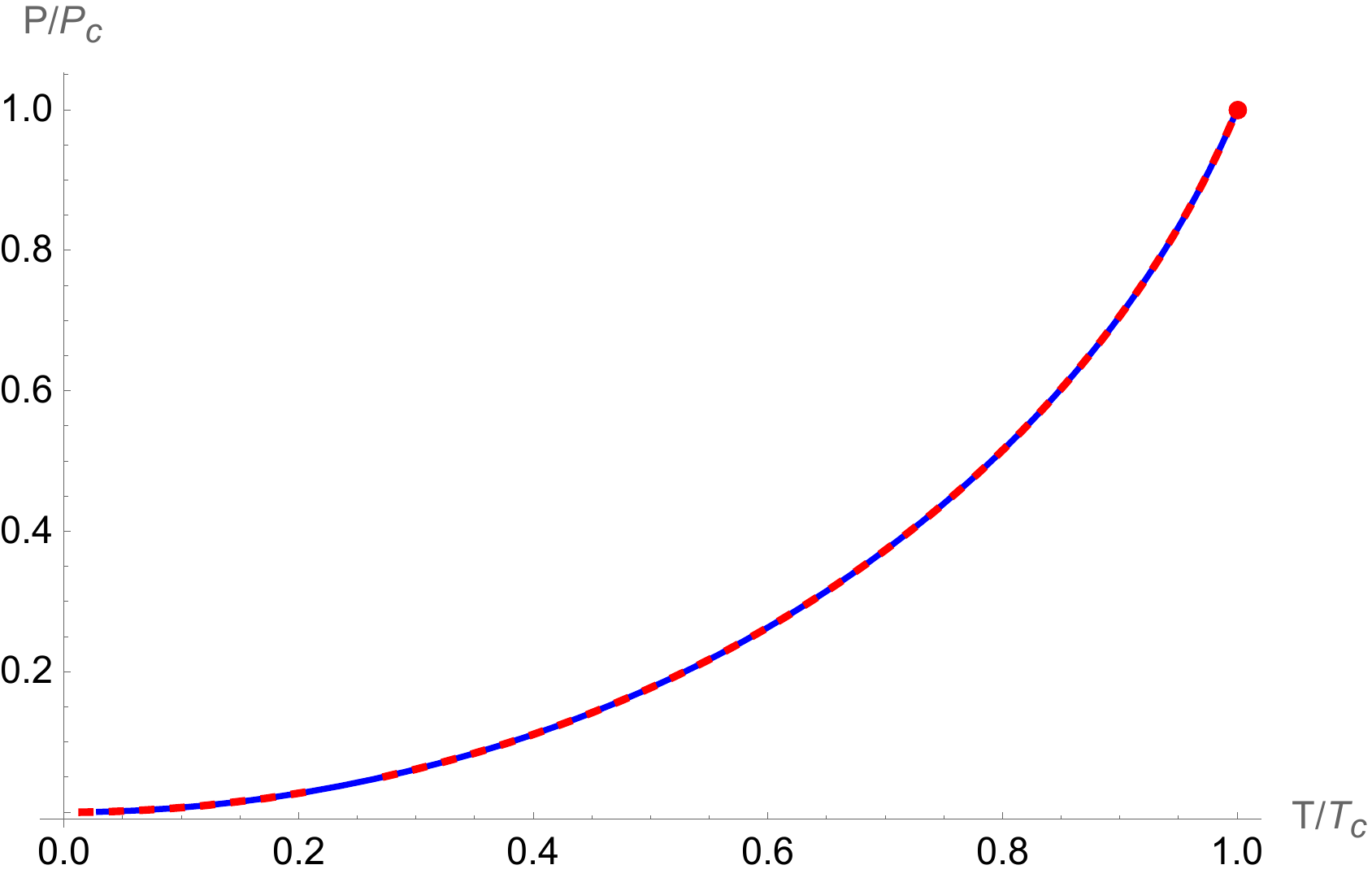}
  \caption{The coexistence curve for the Gauss-Bonnet black hole. The solid line is the analytical solution whereas the dashed line stands for the numerical solution.  }
\label{GBPV}
\end{figure}
Substituting the ansatz (\ref{masteransatz1}) into the equations (\ref{mastereq}) yields
\be p_*=\fft{3(3\phi-1)z_*^2}{z_*^4+(3\phi+1)\phi z_*^2+\phi^2} \,.\ee
Interestingly, nontrivial solution of the function $\phi(z_*)$ turns out to be a simple constant $\phi(z_*)=1$. This leads to
\bea
&& p_*=\fft{6z_*^2}{z_*^4+4z_*^2+1} \,,\nn\\
&& t_*=\fft{z_*(3+p_*z_*^2)}{3z_*^2+1} \,.
\eea
Furthermore, by converting the black hole sizes in terms of the pressure according to
\bea 
&& z_s^2 z_l^2=1\,,  \nn\\
&& z_s^2+z_l^2=\fft{6-4p_*}{p_*}\,,
\eea
and using the fact $t_*=\big( t(z_s)+t(z_l) \big)/2$, we arrive at a compact formula on the $T-P$ plane
\be t_*=\sqrt{\fft{p_*(3-p_*)}{2} } \,.\ee
The result is depicted in Fig. \ref{GBPV} and is perfectly matched with the numerical result reported in \cite{Cai:2013qga}.

\section{quantum BTZ black hole}
Our last example is the $U-\nu$ criticality of the quantum BTZ (qBTZ) black hole studied in \cite{Cui:2025qdy}, where 
\be \nu\equiv \ell/\ell_3 \,,\ee
is a dimensionless parameter characterizing the strength of quantum backreactions of conformal fields in AdS$_3$. Here $\ell_3$ is the bare AdS$_3$ radius and $\ell$ a length parameter inversely proportional to the brane tension, describing the position of the Karch-Randall brane in the classical AdS$_4$ C-metric \cite{Emparan2020}. 

The mass, the temperature and the entropy of the solution are given by \cite{Emparan2020}
\bea
M&=&\fft{\sqrt{1+\nu^2}}{2G_3}\fft{z^2(1-z^3\nu)(1+z\nu)}{(1+3z^2+2z^3\nu)^2}\,,\nn\\
T&=&\fft{z(2+3\nu z+\nu z^3)}{2\pi \ell_3(1+3z^2+2\nu z^3)} \,,\nn\\
S&=&\fft{\pi \ell_3}{G_3}\fft{z\sqrt{1+\nu^2}}{1+3z^2+2\nu z^3}  \,,
\eea
where $G_3$ is the bare Newton's constant in AdS$_3$. Notice that (unlike previous sections) here $z$ is inversely proportional to the event horizon radius $r_h$
\bea
z\equiv\fft{\ell_3}{r_hx_1}\,,
\eea
where $x_1$ is a parameter describing the truncated AdS$_4$ C-metric having a finite black hole in the bulk. Notice that both $z$ and $\nu$ runs in the region $(0\,,+\infty)$.
It is straightforward to test that the ordinary first law $dM=T dS$ holds. 

Previously extended thermodynamics of the solution was partly studied in the literature \cite{Frassino:2022zaz,Frassino:2023wpc,HosseiniMansoori:2024bfi,Frassino:2024bjg} by varying $G_3$ and $\ell_3$. However, it was realized later that emergence of criticality of the solution heavily relies on the quantum backreaction parameter $\nu$. In \cite{Cui:2025qdy}, $\nu$ is treated as a thermodynamic variable directly while $G_3$ and $\ell_3$ are held fixed. In this case, the first law is extended to
\bea
dM=TdS+Ud\nu\,,
\eea
where the chemical potential $U$ reads
\be U=-\fft{z^2\left(\nu+z^4\nu^3+z(\nu^2-1)+z^3(3\nu^2+1)\right)}{2G_3\sqrt{1+\nu^2}(1+3z^2+2z^3\nu)^2}  \,. \ee
Despite that the thermodynamic pressure $P$ and the central charge $C$ are not written explicitly in the extended first law, they actually vary as $\nu$ varies \cite{HosseiniMansoori:2024bfi} and their effects are encoded in the conjugate chemical potential $U$. 

The critical point appears at 
\bea
z_c=1\,,\qquad \nu_c=1\,,\qquad T_c=\fft{1}{2\pi l_3}\,.
\eea
In particular, it was established in \cite{Cui:2025qdy} the small-large black hole transition exists for both $\nu<\nu_c$ ($T<T_c$) and $\nu>\nu_c$ ($T>T_c$).  
\begin{figure}
  \centering
  \includegraphics[width=270pt]{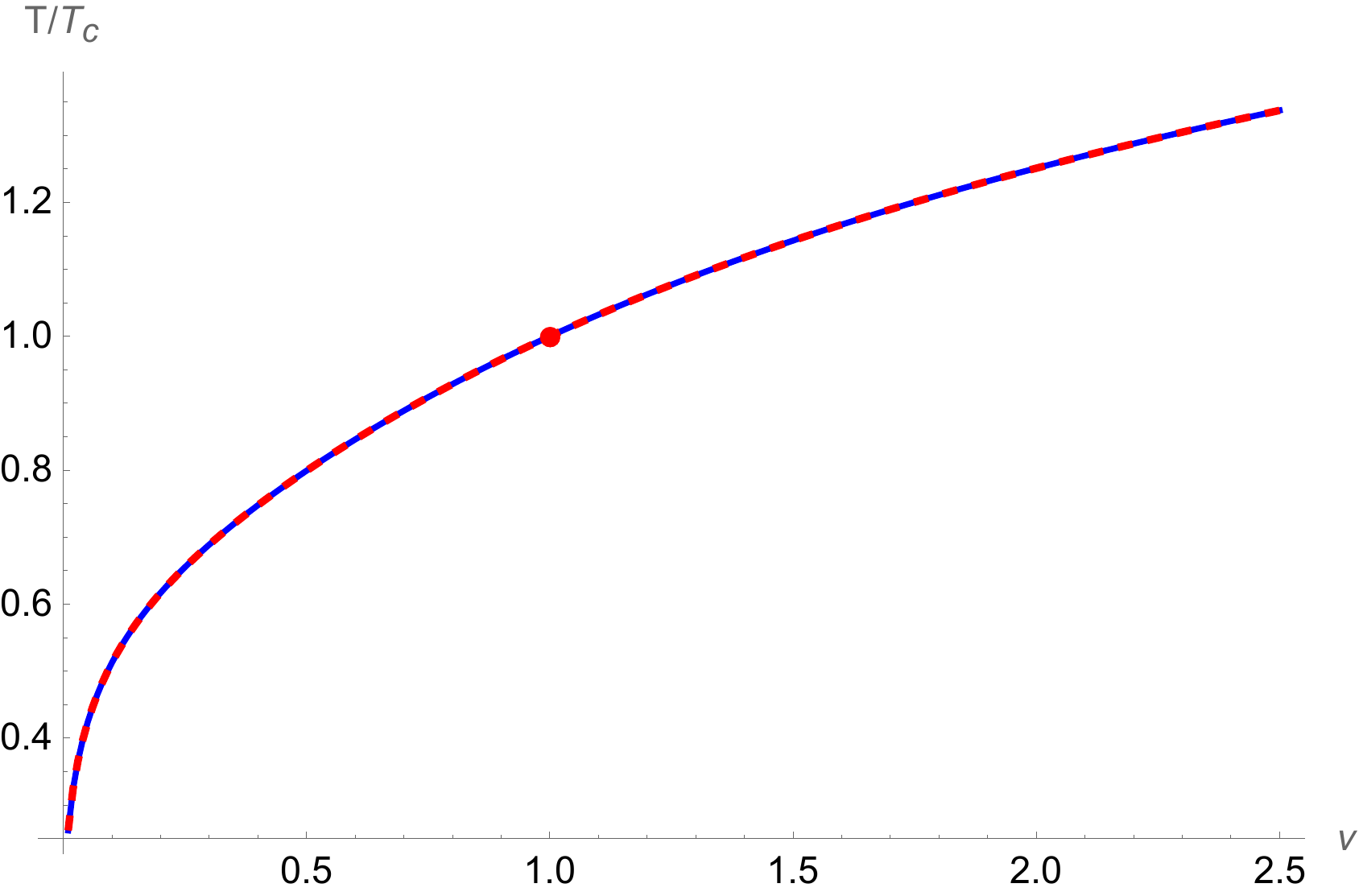}
  \caption{Left: the coexistence curve on the $t-\nu$ plane. The solid line is the analytical solution whereas the dashed line stands for the numerical solution.}
\label{qBTZ}
\end{figure}
To proceed, we work with the normalized temperature and the free energy
 $t=T/T_c\,,f=F/F_c$:
\bea\label{t,s,u}
t&=&\fft{z(2+3z\nu+z^3\nu)}{1+3z^2+2z^3\nu}\,,\nn\\
f&=&\fft{3\sqrt{2}\sqrt{1+\nu^2}z^2\Big(1+z\nu\big(2+z^2(2+z\nu) \big) \Big)}{\Big(1+z^2(3+2z\nu) \Big)^2}\,.
\eea 
Substituting the ansatz (\ref{masteransatz1}) into the equations (\ref{mastereq}) yields
\be 
\nu=\fft{1}{4(3-\phi)\phi^2 z_*^3}\left( z_*^6+3(\phi^2-\phi+1)z_*^4+3\phi( \phi^2-\phi+1 )z_*^2 +\phi^3\pm \sqrt{\Pi}\right)\,,
\ee
where
\bea 
&&\Pi=\left( z_*^4+( 3\phi^2-8\phi+3 )z_*^2+\phi^2 \right)   \nn\\
&&\qquad \times\left(z_*^8+(3\phi^2+2\phi+3)z_*^6+6\phi(\phi^2+3\phi+1)z_*^4+\phi^2(3\phi^2+2\phi+3)z_*^2+\phi^4 \right) \,.
\eea
Despite that the result looks complicated, nontrivial solution to the function $\phi(z_*)$ turns out to be a simple constant $\phi(z_*)=1$ surprisingly. This gives rise to
\bea\label{small solution}
\nu=\fft18\left[(z_*+1/z_*)^3\pm\sqrt{(z_*+1/z_*)^6-64}\right]\,,
\eea
where the ``$\pm$'' sign corresponds to $\nu\geq 1$ and $\nu \leq 1$, respectively. 
 Substituting the result into the equation of state, we derive the coexistence curve on the $t-\nu$ plane (see Fig. \ref{qBTZ})
\be t_*(\nu)=\fft{z_*(\nu)\Big(2+3z_*(\nu)\nu+z_*^3(\nu)\nu \Big)}{1+3z_*^2(\nu)+2z_*^3(\nu)\nu } \,,\ee
where
\be\label{zslnu} z_*(\nu)=\fft12\left( \psi(\nu)\pm\sqrt{\psi(\nu)^2-4} \, \right)\,,\qquad \psi(\nu)=\Big[4(\nu+1/\nu)\Big]^{1/3} \,,\ee
where the sign ``$\pm$'' corresponds to the coexistent small and large black hole respectively. This reproduces the result reported in \cite{Cui:2025qdy}, in which the solution was obtained by a guesswork.  As depicted in Fig. \ref{qBTZ}, it is perfectly matched with the numerical solution. 

\section{Discussions}

In this work, we established a hidden symmetry between the specific volumes of the liquid-gas phases, referred to as self-reciprocal. The property enables us to reduce the coexistence conditions to a single algebraic equation, which in general gives rise to a half-analytical solution to the coexistence line at least. However, for all examples studied in this work, the coexistence line has been solved analytically in terms of a suitable variable $x$ characterizing the liquid-gas phases. In simpler examples, $x$ is simply $z_*$, the normalized specific volume whilst in the other cases $x$ is equal to certain ratio between the specific volumes of the liquid-gas phases. In particular, the master equation for the VdW fluid (charged black hole in higher dimensions) is transcendental (high-power algebraic). These nontrivial examples give us confidence that in general situations the coexistence line might be obtained analytically as well, although whether a suitable variable $x$ can be found depends on details of the system under consideration. We take this as a great advantage of the self-reciprocal property, which deserves further tests in the near future.

\section*{Acknowledgments}

Z.Y. Fan was supported in part by the National Natural Science Foundations of China with Grant No. 11873025.

\appendix

\end{document}